\documentclass{article}
\usepackage{spconf,amsmath,graphicx,hyperref}
\usepackage{booktabs}

\usepackage[utf8]{inputenc}
\usepackage{cite}
\usepackage{amsmath,amssymb,amsfonts}
\usepackage{algorithmic}
\usepackage{graphicx}
\usepackage{textcomp}
\usepackage[table]{xcolor}
\usepackage{booktabs}  
\usepackage[section]{placeins} 
\usepackage{tikz}
\usepackage{multirow}
\usepackage{makecell}
\usepackage{siunitx}
\usepackage{tabularx}
\usepackage{array}
\usepackage{caption}
\usepackage{diagbox}
\usepackage{threeparttable}
\makeatletter
\renewcommand\subsection{\@startsection{subsection}{2}{\z@}%
  {-0.9ex plus -0.2ex minus -0.2ex}
  {0.3ex plus 0.1ex}
  {\normalfont\normalsize\bfseries}}
\renewcommand\subsubsection{\@startsection{subsubsection}{3}{\z@}%
  {-0.7ex plus -0.2ex minus -0.2ex}
  {0.2ex plus 0.1ex}
  {\normalfont\normalsize\itshape}}
\makeatother

\begin{document}
\ninept
%
\title{Beyond Decodability: Do Acoustic Factors Drive Predictions in Speech-Based Alzheimer's Assessment?}
%
\name{
Serli Kopar$^{1,2}$\,\,\,
Alkis Koudounas$^{3}$\,\,\,
Roshan P. Rane$^{1,2}$\,\,\,
Sam Gijsen$^{1,2}$\,\,\,
Paula A. Perez-Toro$^{4}$\,\,\,
Kerstin Ritter$^{1,2}$
}

\address{
$^{1}$ Hertie Institute for AI in Brain Health, Germany 
$^{2}$ Tübingen AI Center, Germany \\
$^{3}$ Sony Group Corporation, Japan 
$^{4}$ Friedrich-Alexander-University Erlangen-Nürnberg, Germany \\
}
%
%
%

%

%
%
%

%
\maketitle
\begin{abstract}
Speech-based Alzheimer's disease (AD) assessments increasingly rely on pretrained self-supervised learning (SSL) models that learn acoustic representations directly from raw audio, exposing the model to recording factors.
We ask whether such factors are merely encoded in SSL representations or can systematically alter predictions. Using ADReSSo and three large SSL backbones, we apply controlled noise and reverberation interventions to participant-speech-only, non-speech, and full-recording audio. We combine layer-wise linear decoding, input- and representation-space interventions, and geometric alignment analysis to distinguish acoustic decodability from influence on AD prediction.
Our results show that controlled acoustic interventions alter AD predictions across all three SSL backbones. Noise, despite showing no significant diagnostic-group difference in the original data, produces the strongest intervention effects. Importantly, these effects are systematically structured relative to the classifier's decision direction, replicate on the held-out test set and reverse when the representation-space intervention direction is reversed. Together, these findings show that high predictive performance and the absence of a significant diagnostic-group difference in a measured acoustic factor are not sufficient for robustness. We argue that intervention-based robustness tests should become standard for trustworthy clinical speech models.
\end{abstract}

\begin{keywords}
self-supervised learning, automatic pathology assessment, Alzheimer’s disease, trustworthy machine learning
\end{keywords}

\section{Introduction}
Speech-based assessments offer a scalable, non-invasive approach to Alzheimer's disease (AD) detection~\cite{fraser2016linguistic,luz2020alzheimers}. Recent systems increasingly rely on self-supervised learning (SSL) models pretrained on large-scale, predominantly non-clinical speech corpora~\cite{baevski2020wav2vec,hsu2021hubert,chen2022wavlm}. Because these models learn directly from raw audio, their representations encode acoustic factors across layers~\cite{pasad2023comparative}, which may provide unintended cues for downstream AD classifiers. Recent work suggests that this risk is not merely theoretical. Gauder et al.~\cite{gauder2024unreliability} and Liu et al.~\cite{liu24f_interspeech} showed that AD can be predicted from non-speech (\textsc{NS}) segments containing silence and background acoustics such as noise. While these findings show that \textsc{NS} regions carry diagnostic information, they leave open whether acoustic factors still influence AD predictions when speech content is available.

In this work, we ask whether acoustic factors are merely decodable from SSL representations or whether they systematically influence downstream AD predictions. This distinction is important because the presence of acoustic information alone does not establish that a classifier functionally relies on it. We therefore move beyond observational decoding and intervene on controlled noise and reverberation both in input space and in the corresponding representation space. By comparing participant-only (\textsc{PAT}), full-recording (\textsc{FULL}), and non-speech (\textsc{NS}) streams, we test whether the classifier sensitivity to these interventions persists when genuine speech content is available. We further analyse the geometry of the induced representation shifts relative to the AD decision direction to test whether intervention effects are directionally structured.

We evaluate this framework on ADReSSo using three large SSL backbones and controlled real and synthetic noise and reverberation degradations. Across models, both input- and representation-space interventions alter AD predictions. Noise produces the strongest changes despite showing no significant diagnostic-group difference in the original recordings. Importantly, the resulting shifts are directionally structured, replicate on the held-out test set, and reverse when the representation-space intervention direction is reversed. Together, these findings show that intervening on acoustic factors systematically shifts downstream AD predictions. Our contributions are threefold: (i) an intervention-based framework that distinguishes acoustic decodability from functional influence on AD predictions; (ii) controlled input- and representation-space interventions showing that acoustic manipulations systematically shift predictions even when speech content is available; and (iii) cross-backbone and held-out validation showing that these effects generalise, are directionally structured and reversible in intervention space. Code is publicly available at \texttt{https://github.com/neselidondurma/beyond-decode}.

\begin{table}[t]
\caption{MMSE and recording characteristics for train and test splits 
}
\label{tab:dataset_statistics_exp_hold}
\centering
\setlength{\tabcolsep}{4pt}
\renewcommand{\arraystretch}{0.84}
\resizebox{\columnwidth}{!}{%
\begin{tabular}{@{}lcccccc@{}}
\toprule
&
\multicolumn{3}{c}{\textbf{Train}} &
\multicolumn{3}{c}{\textbf{Test}} \\
\cmidrule(lr){2-4}
\cmidrule(lr){5-7}
\textbf{Feature}
& \textbf{HC}
& \textbf{AD}
& $\boldsymbol{p}$
& \textbf{HC}
& \textbf{AD}
& $\boldsymbol{p}$ \\
\midrule

$N$
& 79 & 87 & -- & 36 & 35 & -- \\
MMSE
& 29.0\scriptsize{$\pm$1.1}
& 17.4\scriptsize{$\pm$5.3}
& \scriptsize{$<$.001}
& 28.9\scriptsize{$\pm$1.3}
& 18.9\scriptsize{$\pm$5.8}
& \scriptsize{$<$.001} \\

SNR
& 76.2\scriptsize{$\pm$33.5}
& 78.5\scriptsize{$\pm$31.9}
& \scriptsize{.489}
& 76.8\scriptsize{$\pm$33.6}
& 86.9\scriptsize{$\pm$29.6}
& \scriptsize{.073} \\

SRMR
& 5.90\scriptsize{$\pm$2.65}
& 8.10\scriptsize{$\pm$3.88}
& \scriptsize{$<$.001}
& 6.88\scriptsize{$\pm$4.53}
& 5.77\scriptsize{$\pm$3.55}
& \scriptsize{.216} \\

\bottomrule
\end{tabular}
}
\vspace{-3mm}
\end{table}

\section{Methodology}
\subsection{Dataset and Quality Control}
\label{subsec:dataset}
We use the official age- and gender-balanced train ($n=166$) and test ($n=71$) speaker-disjoint splits from the ADReSSo challenge, comprising Cookie Theft picture-description recordings~\cite{luz2021adresso}. Table~\ref{tab:dataset_statistics_exp_hold} reports mean $\pm$ standard deviation of MMSE~\cite{folstein1975mini} and two acoustic factors for healthy controls (HC) and participants with Alzheimer's disease (AD) across train and test splits: WADA-SNR~\cite{kim2008robust}, hereafter SNR, measuring background noise, and SRMR~\cite{falk2010srmr}, measuring reverberation. 
Diagnostic-group differences are assessed using the Mann-Whitney $U$ test~\cite{mann1947test}. 
As expected, diagnostic score MMSE differs significantly by diagnosis in both splits, whereas SNR does not. In contrast, SRMR is significantly higher in AD in the training split ($p<.001$) but not in the test ($p=.216$). 
This split-dependent acoustic structure motivates our controlled interventions.

\begin{figure}[t]
    \centering
    \includegraphics[
        width=0.90\linewidth,
        trim=0mm 3mm 0mm 0mm,
        clip
    ]{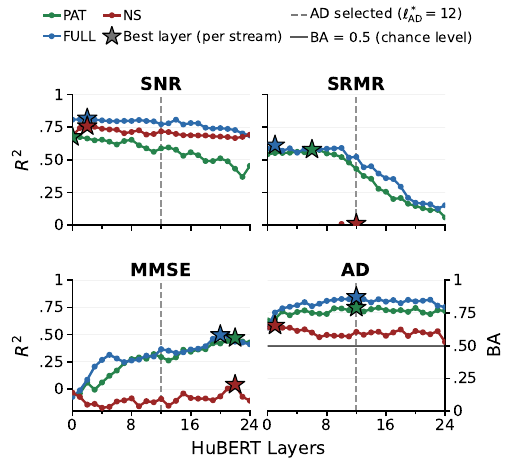}
    \caption{\textbf{E1, linear decoding.}
Layer-wise SNR, SRMR, and MMSE \(R^2\) and AD BA across streams.
}
    \label{fig:e1}
    \vspace{-3mm}
\end{figure}

To examine whether acoustic factor influence differs across recording regions, we construct three streams: \textsc{PAT}, \textsc{FULL}, and \textsc{NS}. Because segmentation errors have been reported~\cite{bang2024alzheimer}, we apply voice activity detection (VAD; \texttt{pyannote.audio v3.4})
~\cite{bredin2021segmentation,bredin2020pyannote}. We define \textsc{PAT} as VAD-positive regions overlapping the provided participant timestamps, \textsc{NS} as VAD-negative regions, and \textsc{FULL} as the complete recording. 
Selected regions are processed separately to avoid artificial acoustic discontinuities~\cite{kraft1993auditory,chappell2002comparison}.

\subsection{Counterfactual Data Generation}
\label{subsec:counterfact}
To assess how controlled interventions on acoustic factors
affect AD predictions, we generate counterfactual recordings in a $2\times2\times4$ design, spanning two acoustic factors (noise/reverberation, as motivated in Sec. \ref{subsec:dataset}), two sources (real (\textsc{R})/ synthetic (\textsc{S})), and four degradation severity levels D1$\rightarrow$D4 (mild$\rightarrow$severe). Noise is introduced additively: \textsc{R} employs environmental recordings from MUSAN~\cite{snyder2015musan}, whereas \textsc{S} applies Gaussian noise at SNR levels of $20, 10, 5,$ and $0$ dB, following established noisy-speech evaluation literature~\cite{hirsch2000aurora}. Reverberation is introduced convolutively using room impulse responses (RIRs): \textsc{R} uses measured RIRs from OpenSLR-28~\cite{ko2017augmentation}, while \textsc{S} applies RIRs simulated with \texttt{pyroomacoustics}~\cite{scheibler2018pyroomacoustics}, targeting $T_{60}$ values of $0.3, 0.5, 0.7$ and $0.9$ s, following and extending the REVERB Challenge settings~\cite{kinoshita2013reverb}. 
For real noise interventions, the same waveform is retained across D1-D4, severity controlled only through amplitude scaling. 
All acoustic degradations are first applied to \textsc{FULL}, after which \textsc{PAT} and \textsc{NS} are reconstructed using the fixed timestamps from Sec.~\ref{subsec:dataset}.



\subsection{Feature Extraction and Prediction Framework}
\label{subsec:pred}
We downmix recordings to mono, resample them to 16~kHz, and extract mean-pooled representations from each layer of three large SSL backbones with distinct pre-training objectives: Wav2Vec2-Large~\cite{baevski2020wav2vec}, HuBERT-Large~\cite{hsu2021hubert}, and WavLM-Large~\cite{chen2022wavlm}. We use Ridge regression for continuous targets (SNR, SRMR and MMSE) and logistic regression for AD classification. Linear models are chosen for their strong performance in ADReSSo~\cite{chen2021automatic} and because they provide an explicit AD-discriminative direction, needed for subsequent analyses.
We evaluate all models using 5-fold nested cross-validation (NCV) on the training set ($n=166$), with feature standardization~\cite{sklearn_standardscaler} and regularization tuning within inner folds. We apply the full experimental framework to all three backbones and present HuBERT in the main results, as it achieves the highest AD performance (BA \(=0.87\)). After fixing all analysis choices from NCV, we retrain on the full training set and evaluate it once on the held-out test set $(n=71)$.



\subsection{Experimental Framework}
We test whether acoustic factors are merely encoded in SSL representations or systematically alter AD predictions through four experiments: (E1) linear decoding, (E2) input-space interventions, (E3) representation-space interventions, and (E4) geometric alignment analysis. Let \(j\) index participants, \(d \in \{\mathrm{noise},\mathrm{reverb}\}\) acoustic degradation factors, \(m \in \{\mathrm{R},\mathrm{S}\}\) real or synthetic degradation sources, \(q \in \{\mathrm{D1,D2,D3,D4}\}\) degradation severity levels, and \(c=(d,m,q)\) the resulting acoustic degradation condition. Further, \(s \in \{\mathrm{PAT},\mathrm{FULL},\mathrm{NS}\}\) indexes streams, \(\ell\) SSL encoder layers, and \(k\) outer folds of NCV. We denote the original \(n=166\) training set by \(\mathcal D\) (Table~\ref{tab:dataset_statistics_exp_hold}). For each \(k\), \(\mathcal D_k\) denotes the held-out data points, while \(\mathcal D_{-k}\) contains the remaining data used for model fitting and inner-cross-validation hyperparameter selection.

\vspace{1mm}
\noindent \textbf{E1, Linear Decoding:}
To characterize where acoustic and clinical information is linearly decodable across SSL layers, we fit 
linear probes to layer-wise representations extracted from the original recordings~\cite{alain2017probes,hewitt2019designing}. For each \((s,\ell,k)\), we probe SNR, SRMR, and MMSE using Ridge regression (reporting \(R^2\)) and AD using logistic regression (reporting balanced accuracy; BA). For subsequent experiments, we freeze the fitted AD classifier and denote its decision function by \(f_{\ell,\mathrm{AD}}^{(s,k)}\), with positive logits indicating AD. For each SSL backbone, the layer with the highest mean outer-fold BA of AD classification is denoted by \(\ell_{\mathrm{AD}}^*\).


\begin{figure*}[t]
    \centering
    \includegraphics[
        width=0.99\textwidth,
        trim=0mm 3mm 0mm 2mm,
        clip
    ]{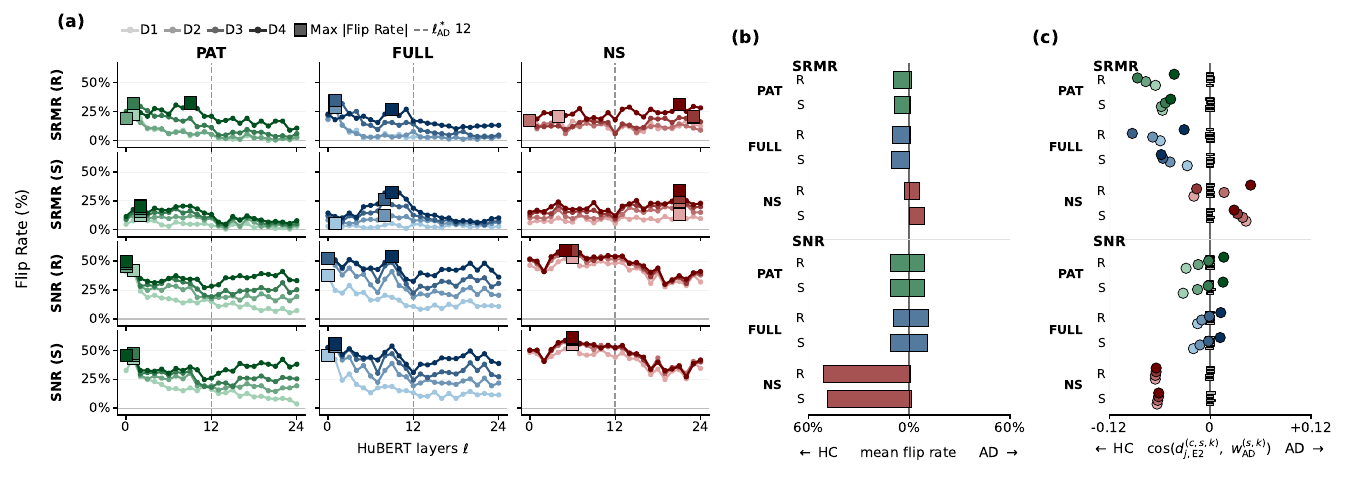}
    \caption{\textbf{E2\&E4, input-space intervention\&alignment.}
\textbf{(a)} AD prediction flip rates across layers for SNR and SRMR interventions; colors denote streams, gradients degradation severity D1--D4, and the dashed line marks \(\ell_{\mathrm{AD}}^*=12\).
\textbf{(b)} HC\(\rightarrow\)AD and AD\(\rightarrow\)HC flips at \(\ell_{\mathrm{AD}}^*=12\), averaged across levels.
\textbf{(c)} Cosine alignment with the AD decision direction at \(\ell_{\mathrm{AD}}^*=12\); gray regions show 95\% random-direction intervals.}
    \label{fig:fig2}
    \vspace{-3mm}
\end{figure*}
\begin{figure*}[t]
    \centering
    \includegraphics[
        width=0.99\textwidth,
        trim=0mm 3mm 0mm 2mm,
        clip
    ]{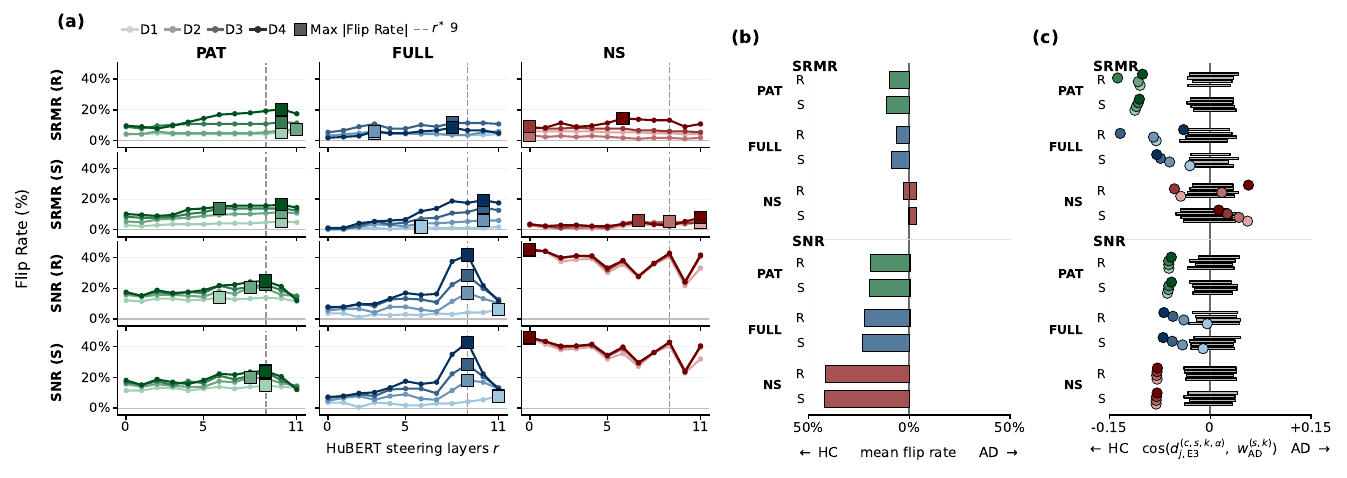}
    \caption{\textbf{E3\&E4, representation-space intervention\&alignment.}
Same panel structure as Fig.~\ref{fig:fig2}; panels b--c show results at \(r^*=9\).}
    \label{fig:fig3}
    \vspace{-3mm}
\end{figure*}
\vspace{1mm}
\noindent \textbf{E2, Input-space Intervention:}
Since linear decodability does not establish that encoded information is used for prediction~\cite{elazar2021amnesic}, we test whether controlled \textit{input space acoustic degradations} alter AD decisions. For each stream \(s\), layer \(\ell\), and outer fold \(k\), we use the frozen AD classifier \(f_{\ell,\mathrm{AD}}^{(s,k)}\) and apply it to the original recording \(x_j\) and of its degraded counterparts \(x_j^{(c)}\), yielding out-of-fold binary predictions \(\hat y_{j,\ell}^{(s)}\) and \(\hat y_{j,\ell}^{(c,s)}\), respectively. We quantify how often acoustic degradation changes the binary AD prediction using the flip rate:
\begin{equation}
\mathrm{FR}_{\ell}^{(c,s)}
=
\frac{1}{|\mathcal D|}
\sum_{j\in\mathcal D}
\mathbb{I}
\!\left[
\hat y_{j,\ell}^{(c,s)}
\neq
\hat y_{j,\ell}^{(s)}
\right].
\label{eq:e2}
\end{equation}
We further decompose the flip rate into \(\mathrm{HC}\rightarrow\mathrm{AD}\) and \(\mathrm{AD}\rightarrow\mathrm{HC}\) transitions, such that their sum equals the total flip rate $\mathrm{FR}_{\ell}^{(c,s)}$.

\vspace{1mm}
\noindent \textbf{E3, Representation-space Intervention:}
To test whether controlled \textit{representation-space acoustic degradations} alter AD decisions, we steer hidden representations at layer \(r\), propagate the steered representation through the remaining encoder layers to \(\ell_{\mathrm{AD}}^*\), and evaluate the resulting change using the same frozen AD classifier \(f_{\ell^*,\mathrm{AD}}^{(s,k)}\).  More formally, let \(h_r^{(s)}(x)\) denote the hidden representation of input \(x\) from stream \(s\) at layer \(r\). For each degradation condition \(c\), stream \(s\), and participant \(j\), we first compute the degradation shift as:
\begin{equation}
\underbrace{\delta_{r,j}^{(c,s)}}_{\text{degradation shift}}
=
\underbrace{h_r^{(s)}(x_j^{(c)})}_{\text{degraded representation}}
-
\underbrace{h_r^{(s)}(x_j)}_{\text{clean representation}}.
\label{eq:e3_shift}
\end{equation}
We average these degradation shifts over \(\mathcal D_{-k}\) and normalize the result to obtain a shared acoustic degradation direction \(u_r^{(c,s,k)}\), following the general activation-difference averaging principle also used in Contrastive Activation Addition~\cite{rimsky2024steering}. We then project each \(\mathcal D_{-k}\) degradation shift onto this shared direction and use the median projection magnitude, \(\rho_r^{(c,s,k)}\), as a robust estimate of the typical degradation strength.
 For each held-out participant \(j\in\mathcal D_k\), we calculate the \textit{steered representation} as:
\begin{equation}
\underbrace{\tilde h_{r,j}^{(c,s,k,\alpha)}}_{\text{steered representation}}
=
\underbrace{h_r^{(s)}(x_j)}_{\text{clean representation}}
+
\underbrace{\alpha\,\rho_r^{(c,s,k)}u_r^{(c,s,k)}}_{\text{acoustic steering shift}}.
\label{eq:e3_steer}
\end{equation}
where \(\alpha\) controls the steering strength and direction. We propagate the \textit{steered
representation} \(\tilde h_{r,j}^{(c,s,k,\alpha)}\) through
the remaining frozen SSL encoder layers up to \(\ell_{\mathrm{AD}}^*\),
yielding the \textit{propagated representation} \(\tilde h_{\ell_{\mathrm{AD}}^* \leftarrow r,\, j}^{\,(c,s,k,\alpha)}\).
We quantify steering-induced decision changes using Eq.~\ref{eq:e2} and same frozen classifier  \(f_{\ell^*,\mathrm{AD}}^{(s,k)}\). Similar to E2, we further decompose the flip rate into \(\mathrm{HC}\rightarrow\mathrm{AD}\) and \(\mathrm{AD}\rightarrow\mathrm{HC}\). We report \(\alpha=+1\), while \(\alpha=-1\) serves as a reversibility control. \(r^*\) is selected on the training set as the layer with the highest flip rate.


\vspace{1mm}
\noindent \textbf{E4, Geometric alignment with the AD decision direction:}
To geometrically characterize the E2 and E3 effects, we measure how their intervention-induced displacements align with the frozen classifier's AD decision direction using cosine similarity \(A\)~\cite{cosinsim}. The E2 and E3 displacements are calculated as:
\begin{equation}
\begin{aligned}
d_{j,\mathrm{E2}}^{(c,s,k)}
&=
\underbrace{h_{\ell_{\mathrm{AD}}^*}^{(s)}(x_j^{(c)})}_{\text{degraded representation}}
-
\underbrace{h_{\ell_{\mathrm{AD}}^*}^{(s)}(x_j)}_{\text{clean representation}},
\\
d_{j,r,\mathrm{E3}}^{(c,s,k,\alpha)}
&=
\underbrace{\tilde h_{\ell_{\mathrm{AD}}^* \leftarrow r,\, j}^{\,(c,s,k,\alpha)}}_{\text{propagated representation}}
-
\underbrace{h_{\ell_{\mathrm{AD}}^*}^{(s)}(x_j)}_{\text{clean representation}}.
\end{aligned}
\label{eq:e4_displacements}
\end{equation}
Because the frozen classifier operates on standardized features, we transform each displacement into the same feature space as \(\widehat d=d\oslash\sigma^{(s,k)}\), where \(\sigma^{(s,k)}\) is the scaler of \(f_{\ell^*,\mathrm{AD}}^{(s,k)}\). We then compute the cosine similarity between \(\widehat d\) and the frozen classifier weight vector \(w_{\ell^*,\mathrm{AD}}^{(s,k)}\). Positive \(A\) denotes displacement toward the AD decision direction. Notably, for the linear classifier, \(w^\top\widehat{d}\) determines the logit change, so \(A\) serves as a geometric characterization of the intervention effect rather than independent evidence. 

\vspace{1mm}
\noindent \textbf{Random Direction Controls:}
To assess the E4 alignment strength, we generate 100 random-orientation control vectors.
We sample isotropic directions magnitude-matched to the E2 displacement \(d_{j,\mathrm{E2}}^{(c,s,k)}\) and the E3 acoustic steering shift \(\alpha\rho_r^{(c,s,k)}u_r^{(c,s,k)}\), respectively. E3 controls are propagated through the remaining SSL layers as in the original intervention. The central 95\% of the random-control cosine similarities define the null intervals shown in Figs.~\ref{fig:fig2}c and~\ref{fig:fig3}c as gray, y-axis-centered bands.



\begin{minipage}{0.8\columnwidth}
    \centering
    \includegraphics[
        width=\linewidth,
        trim=0mm 1.5mm 0mm 0mm,
        clip
    ]{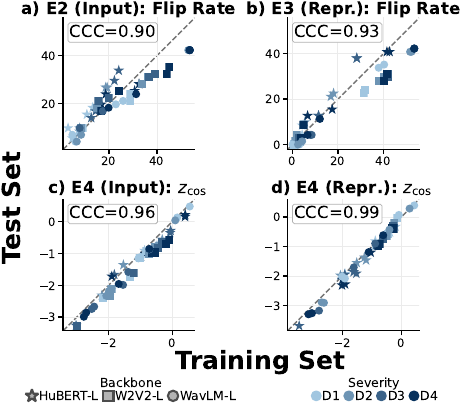}
    \captionof{figure}{Backbone and Held-Out-Test Generalisation}
    \label{fig:gen}
    \vspace{-3mm}
\end{minipage}

\section{Results}
\noindent \textbf{E1, Linear Decoding:} We first probe how acoustic and clinical information is distributed across HuBERT layers. In Fig.~\ref{fig:e1}, acoustic factors show distinct depth profiles. SNR remains highly linearly decodable across layers and streams, reaching \(R^2\approx0.8\) in \textsc{FULL}, whereas SRMR is decodable in early layers of the speech-containing \textsc{PAT} and \textsc{FULL} streams (\(R^2\approx0.6\)) and declines markedly with depth. In contrast, MMSE decodability in \textsc{PAT} and \textsc{FULL} increases across the layers, reaching \(R^2\approx0.5\) in later layers. AD classification performance is strongest in intermediate layers, peaking at layer~12 (\(\ell_{\mathrm{AD}}^*\)). Notably, \textsc{NS} retains strong SNR information and above-chance AD classification, while SRMR and MMSE are essentially not linearly decodable, consistent with SRMR's reliance on speech modulation structure that is largely absent from NS.

\vspace{1mm}
\noindent \textbf{E2 \& E4, Input-space Intervention \& Alignment:}
We next test whether controlled acoustic interventions alter AD predictions across HuBERT layers. Fig.~\ref{fig:fig2}a shows that SNR degradation causes substantially more prediction flips than SRMR when clean and degraded inputs are evaluated with the same frozen layer-wise classifier \(f_{\ell,\mathrm{AD}}^{(s,k)}\). The largest difference occurs in \textsc{FULL} at \(\ell=1\), with flip rates of \(51\%\) for SNR(S) versus \(8\%\) for SRMR(S). Flip rates generally increase with degradation severity (D1\(\rightarrow\)D4). \textsc{NS} under SNR is the main exception, with even D1 producing flip rates comparable to D4, indicating high SNR sensitivity of non-speech regions. Fig.~\ref{fig:fig2}b shows the direction of these flips, averaged across D1--D4. Under SRMR interventions, AD\(\rightarrow\)HC flips dominate in \textsc{PAT} and \textsc{FULL}, whereas HC\(\rightarrow\)AD flips dominate in \textsc{NS}. Under SNR interventions, the clearest asymmetry occurs in \textsc{NS}, where AD\(\rightarrow\)HC flips dominate. Fig.~\ref{fig:fig2}c provides the corresponding geometric view at \(\ell_{\mathrm{AD}}^*\), with positive cosine similarity indicating alignment toward AD. Overall, 44/48 conditions (91.7\%) fall outside the central 95\% range of \(n=100\) magnitude-matched random directions. As expected for a linear classifier, the alignment patterns correspond closely to the directional flips: under SRMR, \textsc{PAT} and \textsc{FULL} shift mainly toward HC while \textsc{NS} shifts toward AD. Under SNR, \textsc{NS} remains strongly HC-directed, whereas \textsc{PAT} and \textsc{FULL} show a severity-dependent shift toward AD as degradation increases.

\vspace{1mm}
\noindent \textbf{E3 \& E4, Representation-space Intervention \& Alignment:}
We next test whether the input-space sensitivity observed in E2 also appears when acoustic degradation directions are introduced directly in representation space. Fig.~\ref{fig:fig3}a shows clear layer-dependent steering effects, with consistently higher flip rates for SNR than SRMR. The largest effect occurs at \(r^*=9\), where steering along the D4 SNR-derived direction in \textsc{FULL} flips \(42\%\) of predictions. This peak is not explained by perturbation magnitude, since \(\rho_r\) decreases with depth, from \(83\) at \(r=0\) to \(64\) at \(r=9\). Fig.~\ref{fig:fig3}b shows that these flips are predominantly AD\(\rightarrow\)HC in the speech-containing \textsc{PAT} and \textsc{FULL} streams for both SNR- and SRMR-derived directions. \textsc{NS} differs: SNR steering remains HC-directed, whereas SRMR produces fewer, predominantly AD-directed flips. Similarly, Fig.~\ref{fig:fig3}c shows that \textsc{PAT} and \textsc{FULL} shifts are mainly aligned toward HC and often fall outside the central 95\% range of \(n=100\) magnitude-matched random directions. Finally, reversing the steering coefficient from \(\alpha=+1\) to \(-1\) reverses the logit effect (\(r_{\mathrm{Pearson}}=-0.98\)), demonstrating strong directional reversibility.


\vspace{1mm}
\noindent \textbf{Backbone and Dataset Generalisation:}
\label{subsec:gentest}
Finally, we assess generalisation across SSL backbones and the held-out test set, with all analysis choices fixed from the train set (Sec.~\ref{subsec:pred}). Fig.~\ref{fig:gen} compares train- and test-set effects using Lin's concordance correlation coefficient (CCC)~\cite{lin1989concordance}, which captures agreement in both association and magnitude. In Fig.~\ref{fig:gen}c--d, cosine alignment is null-standardised as \(z_{\cos}=(A_{\mathrm{obs}}-\mu(A_{\mathrm{rand}}))/\sigma(A_{\mathrm{rand}})\) using \(n=100\) magnitude-matched random directions, with negative values indicating HC-directed alignment. Overall, train and test effects show strong agreement, with CCC \(=0.90\!-\!0.99\), \(r_{\mathrm{Pearson}}=0.95\!-\!0.99\), and MAE \(=0.050\!-\!0.236\) across experiments.

\section{Discussion and Limitations}
A common assumption in speech-based clinical AI is that high predictive performance and no significant group differences in measured acoustic factors imply robust predictions. Our results challenge this assumption directly: acoustic interventions alter AD predictions in both input and representation space, including for SNR despite no significant group difference. This extends prior decodability~\cite{pasad2021layerwise} and non-speech-only studies~\cite{liu24f_interspeech,gauder2024unreliability,jouaiti2026noise} by showing that interventions on acoustic factors influence predictions even when genuine speech content is available. E4 further shows that these intervention effects are not captured by a single direction: alignment varies across factors, streams, and severities, and input- and representation-space interventions can shift decisions differently. 
This heterogeneity argues against a simple fix based on identifying and projecting out a single global ‘acoustic confound direction,’ and instead suggests a more distributed sensitivity.


These findings have a direct practical implication: The absence of a statistically significant group difference in an acoustic factor is not sufficient evidence of robustness. In our data, SNR showed no significant group difference, yet produced the largest intervention-induced flips. Clinical speech models should therefore be evaluated with intervention-based robustness alongside predictive accuracy.


Our study has some limitations that motivate future work. We examine only noise and reverberation, chosen for their established role in speech processing~\cite{kinoshita2016reverb,li2014overview} and links to AD prediction~\cite{novikova2022impact}.  Our interventions, while spanning real and synthetic sources, are still applied post-hoc. Validating these with in-the-wild multilingual recordings would clarify generalisability of the effects.

Overall, predictive accuracy and acoustic decodability alone are insufficient for trustworthiness. Intervention-based audits are needed to test whether predictions remain stable under acoustic variation that is clinically irrelevant but practically unavoidable.

\vfill\pagebreak
\section{Compliance with Ethical Standards}
This study uses existing ADReSSo data derived from the DementiaBank, collected with informed consent. All data were used in accordance with DementiaBank access requirements.

\bibliographystyle{IEEEbib}
\bibliography{strings,refs}

\end{document}